\documentclass[aps,pre,reprint,superscriptaddress]{revtex4-1}

\usepackage{amsmath}
\usepackage{txfonts}

\usepackage[T1]{fontenc}
\usepackage{textcomp}
\usepackage[utf8]{inputenc}

\usepackage[pdftex]{graphicx}

\usepackage{enumerate}

\usepackage{bm}
\bmdefine{\aVector}{a}
\bmdefine{\AVector}{A}
\bmdefine{\bVector}{b}
\bmdefine{\BVector}{B}
\bmdefine{\cVector}{c}
\bmdefine{\CVector}{C}
\bmdefine{\eVector}{e}
\bmdefine{\EVector}{E}
\bmdefine{\fVector}{f}
\bmdefine{\FVector}{F}
\bmdefine{\gVector}{g}
\bmdefine{\pVector}{p}
\bmdefine{\PVector}{P}
\bmdefine{\qVector}{q}
\bmdefine{\QVector}{Q}
\bmdefine{\rVector}{r}
\bmdefine{\RVector}{R}
\bmdefine{\sVector}{s}
\bmdefine{\uVector}{u}
\bmdefine{\vVector}{v}
\bmdefine{\VVector}{V}
\bmdefine{\muVector}{\mu}
\bmdefine{\OmegaVector}{\Omega}

\begin{document}


\title{Conducting transition analysis of thin films composed of long flexible macromolecules: Percolation study}


\author{Yuki Norizoe}
\email[E-mail: ]{norizoe@cmpt.phys.tohoku.ac.jp}
\altaffiliation[Present address: ]{Department of Physics, Tohoku University, 980-8578 Sendai, Japan.}
\affiliation{Technology Research Association for Single Wall Carbon Nanotubes (TASC) - Central 2-1, 1-1-1 Umezono, Tsukuba, Ibaraki 305-8568, Japan}
\affiliation{National Institute of Advanced Industrial Science and Technology (AIST) - Central 2-1, 1-1-1 Umezono, Tsukuba, Ibaraki 305-8568, Japan}

\author{Hiroshi Morita}
\affiliation{National Institute of Advanced Industrial Science and Technology (AIST) - Central 2-1, 1-1-1 Umezono, Tsukuba, Ibaraki 305-8568, Japan}


\date{July 30, 2019}

\begin{abstract}
Simulating percolation and critical phenomena of labelled species inside films composed of single-component linear homogeneous macromolecules using molecular Monte Carlo method in 3 dimensions, we study dependence of these conducting transition and critical phenomena upon both thermal movement, i.e. spontaneous mobility, and extra-molecular topological constraints of the molecules. Systems containing topological constraints and/or composed of immobile particles, e.g. lattice models and chemical gelation, were studied in conventional works on percolation. Coordinates of the randomly distributed particles in the conventional lattice models are limited to discrete lattice points. Moreover, each particle is spatially fixed at the distributed position, which results in a temporally unchanged network structure. Although each polymer in the chemical gels can spontaneously move in the continuous space, the network structure is fixed when cross-linking reaction ends. By contrast to these conventional systems, all the molecules in the present system freely move and spontaneously diffuse in the continuous space. The network structure of the present molecules continues changing dynamically. The percolation and critical phenomena of such dynamic network structures are examined here. We reveal that these phenomena also occur in the present system, and that both the universality class and percolation threshold are independent from the extra-molecular topological constraints.
\end{abstract}


\maketitle

\section{Introduction}
\label{sec:Introduction}
Molecular systems, such as films and solutions, composed of spontaneously-moving (mobile) macromolecules, \textit{e.g.} polymers and carbon nanotubes, have been widely studied in soft matter science, \textit{e.g.} physics and chemistry~\cite{Hur:2015,Marencic:2012,Lefevre:2010}, and extensively utilized in nanotechnology and industrial applications, \textit{e.g.} coating and nanolithography~\cite{Gu:2014,Lefevre:2010,Bosworth:2008}. Physical properties and collective phenomena of the molecular systems, such as electrical conduction, elasticity, viscosity, \textit{etc.} depend on network structures, which always continue to spontaneously change, among the mobile molecules. The understanding of these dynamically changing network structures enhances that of the physical properties of the molecular systems. Percolation theory could help and facilitate such understanding of the network structures. For example, the percolation threshold and conductive network structure among molecules, which are discussed in general percolation theory, are directly linked to design and development of conductive or non-conductive materials, so that the understanding of percolation theory leads to improvements to such material design.

General percolation theory is known as a scientific field of collective phenomena resulting from connections and links. Percolation phenomena occur in and are related to a variety of physical systems and phenomena, \textit{e.g.} electrical conduction, mercury porosimetry, forest fire, polymer gels, and string-like colloidal assembly~\cite{Stauffer1985,Sahimi:ApplicationsOfPercolationTheory,Norizoe:2012JCP,Norizoe:2005}. General percolation theory indicates that fractal structure of large percolation clusters, divergence of various physical quantities, power law of cluster size distribution, and other critical phenomena also occur at the percolation transition. However, in conventional works on percolation theory, such percolation and concomitant critical phenomena were studied for systems containing topological constraints and/or consisting of immobile particles such as lattice models~\cite{Stauffer1985} and network structure of chemical gels~\cite{Sahimi:ApplicationsOfPercolationTheory}. The network structures of such conventional systems are temporally fixed. For example, the percolation transition of particles, which are randomly distributed one by one over lattices, was typically discussed in the conventional works~\cite{Stauffer1985}. The spatial coordinates of each particle in this conventional lattice model are limited to the discrete lattice points and fixed at the initially distributed position. The network structure among these particles is also fixed when the distribution of the particles stops. Such constraints also apply to the systems of the chemical gels. Although polymers composing the gels can spontaneously move in the continuous space, cross-links between the polymers are never disconnected. The network structure of the connected polymers is fixed when the cross-linking reaction ends~\cite{Sahimi:ApplicationsOfPercolationTheory}. In other words, the percolation phenomena of static networks were studied in such conventional works.

By contrast, here we simulate percolation phenomena of dynamic networks, in which the network structure always continues changing spontaneously, for the purpose of overcoming the above limitations of the conventional percolation theory, and enhancing the understanding of both the percolation theory and universal phase behaviour of molecular systems. We demonstrate that the percolation and critical phenomena are also observed in the dynamic networks. Moreover, comparing simulation results, such as the percolation threshold and critical exponents, between systems with and without the above extra-molecular topological constraints, we examine whether the percolation phenomena of the network among the molecules are dependent on the constraints.

The chemical gels and bulk polymer solutions could be simulated and compared for the above purpose. However, the system of the gels includes additional unique simulation parameters related to the cross-linking reaction, such as reaction rate and reaction time. These unique properties significantly complicate the equal condition in which the gel and bulk solution are compared. Therefore, here we choose other systems instead of the gel and bulk solution.

Homopolymer brushes~\cite{Norizoe:2014JCP}, \textit{i.e.} single-component monodisperse linear flexible homopolymers grafted onto a planar substrate, are also known as a typical molecular system with the above constraint. Molecular films (thin fluid layers) are similar to the brushes, and free from this extra-molecular constraint, \textit{i.e.} grafting. When the thickness, density, and other parameters of the brush and film are fixed at the equal values, the molecules inside the film can freely move and diffuse inside the same spatial region of the film as the region of the brush. These systems could be simulated and compared for our purpose.


Thus, in the present work, simulating the film (fluid membrane) composed of single-component monodisperse linear flexible homogeneous molecules placed on a planar hard substrate, we study the percolation and critical phenomena of dynamic networks without the above constraints. The network structure of the molecules spontaneously and temporally changes in the three-dimensional continuous space. These characteristics illustrate that the present system is, to determine the universal phase behaviour of percolation systems, more basic and general than the above conventional systems are. Furthermore, the present film is also more basic and general than those composed of inhomogeneous molecules. Such sophisticated films have recently attracted broad attention, and exhibit micro domain patterns and other concomitant various phenomena according to the chemical and physical properties of the chemically immiscible blocks along each molecule. However, such unique phenomena resulting from the unique chemical and physical properties of the sophisticated films disturb studies about universal physical phenomena occurring in arbitrary films. The simulation of the present basic film could result in a study on the universal phase behaviour found in arbitrary films.

Here we determine the percolation threshold (transition density) and study concomitant critical phenomena. Values of critical exponents and fractal dimension of the percolation clusters are also determined. These values are compared with those of the homopolymer brushes. The system of these brushes is the same as the present system except that the polymers of the brushes are grafted onto the hard planar substrate~\cite{Norizoe:2013EPL,Norizoe:2014JCP}, so that the dependence of the phase behaviour on the topological constraints, \textit{i.e.} grafting, is obtained.

Percolation phenomena of macromolecular systems were also studied in early works in two dimensions using lattice models~\cite{Kondrat:2002,Adamczyk:2009,Zerko:2012,Polanowski:2013} and off-lattice models~\cite{Yethiraj:2003}.
The universality of the polymer size was also observed in 2-dimensional lattice models recently~\cite{Polanowski:2018}.

In the present work, we construct a model system and run the simulation in a coarse-grained scale in the continuous 3-dimensional space. This coarse-grained model is discussed in sections~\ref{sec:ConceptOfTheSimulationModel} and \ref{sec:Solvent-freeModel}. The simulation method is given in section~\ref{sec:SimulationMethods}. How to construct the dense film on the planar hard substrate is also illustrated in sections~\ref{sec:Solvent-freeModel} and \ref{sec:SimulationMethods}. Simulation results are shown in section~\ref{sec:SimulationResults}. Finally, we summarize the present manuscript in section~\ref{sec:Conclusions}.

A list of symbols used in the present work is shown in Table~\ref{tab:ListOfSymbols} for easy reference.
\begin{table*}[!htb]
	\caption[]{List of symbols used in the present work.}
	\label{tab:ListOfSymbols}
	\centering
	\begin{tabular}{cl} \hline
		Symbol (and value, if fixed) & \qquad Definition and notes  \\
		\hline
		$N = 32$ & Degrees of polymerization  \\
		$V$ & System volume  \\
		$k_B T$ & Thermal energy, chosen as the unit energy  \\
		$H_\text{spring}$ & Harmonic spring potential, defined in eq.~\eqref{eq:HarmonicSpringPotential}  \\
		$k_\text{spring} = \left. 3 ( N - 1 ) \middle/ R_e^2 \right.$ & Spring constant  \\
		$R_e = b \sqrt{N}$ & Root mean square of the end to end distance of an ideal molecular chain, unit length  \\
		$b$ & Average bond length between bonded segments of ideal molecular chains  \\
		$H_{\text{non-bonded}}$ & Non-bonded interaction potential defined in eq.~\eqref{eq:NonidealFreeEnergyDimensionless}  \\
		$\rho_\beta ( \rVector )$ & Local volumetric number density of $\beta$-segments at the spatial coordinate $\rVector$  \\
		$\rho_\text{p(label)}' ( \rVector )$ & Dimensionless local molecular density of the labelled molecules, defined in eq.~\eqref{eq:DimensionlessPolymerDensity}  \\
		$\rho_\text{p(unlabel)}' ( \rVector )$ & Dimensionless local molecular density of the unlabelled molecules, defined in eq.~\eqref{eq:DimensionlessPolymerDensity}  \\
		$v' = 7.0 v'_c$ & Dimensionless attractive interaction strength among the segments  \\
		$w' = 1.0$ & Dimensionless repulsive interaction strength among the segments  \\
		$\rho_p' = \left. n_p R_e^3 \middle/ V \right.$ & Dimensionless average volumetric molecular density of the single-component system  \\
		$n_p$ & Number of molecules in the single-component system  \\
		$\left( \rho_{pc}' = \left. 1 \middle/ \sqrt{ 2w' } \right. \, , \, v'_c = 2 \sqrt{2w'} \, \right)$ & Critical point of the binodal line in $\rho_p' v'$-plane, based on the mean-field theory  \\
		$\varDelta L = (1/6) R_e$ & Grid spacing  \\
		$\left( L_x, L_y, L_z = 6.0 R_e \right)$ & System size  \\
		$n_S^\text{(total)}$ & Total number of segments in the system  \\
		$\rho_\text{p(label)} = \left. n_\text{p(label)} \middle/ ( L_x L_y \times R_e ) \right.$ & Average volumetric density of the labelled molecules in the region of the layer  \\
		$\rho_\text{p(label)}' = \rho_\text{p(label)} R_e^3$ & Dimensionless density reduced from $\rho_\text{p(label)}$  \\
		$n_\text{p(label)}$ & Number of labelled molecules  \\
		$n_\text{p(unlabel)}$ & Number of unlabelled molecules  \\
		$n_p^\text{(total)} = n_\text{p(label)} + n_\text{p(unlabel)}$ & Total number of labelled and unlabelled molecules in the system  \\
		$\rho_p^\text{(total)} = \left. n_p^\text{(total)} \middle/ V \right.$ & Average total molecular density in the system  \\
		$n(s) = m(s) / N_\text{cell}$ & Cluster size distribution where $s$ denotes the cluster size  \\
		$m(s)$ & Number of clusters with the size $s$ found in the system  \\
		$N_\text{cell} = \left. L_x L_y \middle/ ( \varDelta L )^2 \right.$ & Total number of the cells on the substrate  \\
		$\tau$ & Critical exponent, referred to as Fisher exponent  \\
		$D_f$ & Fractal dimension of the percolation clusters in the vicinity of the transition point  \\
		$n_\text{sub} = N / N_\text{sub}$ & Number of subchains in one molecule  \\
		$N_\text{sub}$ & Number of segments in one subchain  \\
		\hline
	\end{tabular}
\end{table*}

\section{Concept of the simulation model}
\label{sec:ConceptOfTheSimulationModel}
As was discussed in section~\ref{sec:Introduction}, the film is simulated in the present work. The film itself always bridges both the edges of the substrate, and is in the percolation phase. Here we mix labelled molecules into the film. The percolation transition of the labelled molecules occurs according to the ratio of the labelled molecules to the whole film. We study the percolation and concomitant critical phenomena of the network of these labelled molecules inside the film.

When all the molecules in a dense film on a planar hard substrate are labelled, the labelled molecules cover the whole substrate and bridge all the edges of the substrate. In contrast, when the fraction of the labelled molecules to the film is extremely low, the labelled molecules are isolated from each other. This indicates that the percolation transition of the labelled molecules in the film occurs at the intermediate density of the labelled molecules. According to the general percolation theory~\cite{Stauffer1985}, fractal structure of this percolation cluster of the labelled molecules is observed at the percolation transition. Both the labelled and unlabelled molecules freely move and diffuse in the continuous space in the present work.

\section{Solvent-free model}
\label{sec:Solvent-freeModel}
In the present work, we simulate the dense films lying on a planar hard substrate, which stays at the bottom of the system box. Spatial regions inside the system box above this molecular layer are filled with a huge number of solvent particles. These particles cause drastic increase in the amount of computation and significantly retard the simulation although we intend to simulate the molecular layer itself rather than the solvents. Therefore, here a binary solvent-free coarse-grained model~\cite{Drouffe:1991,Daoulas:2010} of monodisperse linear flexible molecules proposed by M{\"{u}}ller and Daoulas~\cite{DoctoralThesis,Norizoe:2010Faraday} is employed, where these two molecular species denote the labelled and unlabelled molecules respectively. This 3-dimensional (3-D) model was also utilized in our recent works~\cite{Norizoe:2010Faraday,Norizoe:2013EPL,Norizoe:2014JCP}, so that the present simulation results can be compared with the ones of these recent works, especially the results of the brush~\cite{Norizoe:2014JCP}, in the equal condition when the simulation conditions are set at the equal values. The validity of the recent work~\cite{Norizoe:2013EPL} using the solvent-free model is also confirmed in another experimental work~\cite{Murakami:2016}. This model is briefly summarized here.

In the solvent-free model, explicit degrees of freedom of solvents are integrated out. The solvents are replaced with an effective non-bonded interaction between solute molecules. This coarse-graining significantly diminishes degrees of freedom of the system and computational time demanded for simulation of the film lying in 3-D space. Each molecule of both the molecular species is composed of $N$ coarse-grained segments. These $N$ segments are linearly connected by harmonic spring potential, $H_\text{spring}$,
\begin{equation}
	\label{eq:HarmonicSpringPotential}
	\left. H_\text{spring} \middle/ k_B T \right. = \left( k_\text{spring} \middle/ 2 \right) r^2,
\end{equation}
where $k_B T$, which is chosen as the unit energy, denotes the thermal energy, $r$ is the distance between the centres of the pair of the connected coarse-grained segments, and the spring constant is fixed at $k_\text{spring} = 3 ( N - 1 ) / R_e^2$. $R_e$, which is chosen as the unit length, represents the root mean square of the end to end distance of an ideal molecular chain with the same molecular architecture. Equation~\eqref{eq:HarmonicSpringPotential} is corresponding to the bonded interaction between the coarse-grained segments. S(label) and S(unlabel) denote the segment species of the labelled and unlabelled molecular species, respectively. The non-bonded interaction potential of the present solvent-free model, denoted by $H_{\text{non-bonded}}$, is defined as a functional,
\begin{align}
	\label{eq:NonidealFreeEnergyDimensionless}
	\frac{ H_{\text{non-bonded}} }{k_B T} = \int_V \frac{dV}{{R_e}^3} & \left( -\frac{1}{2}v' \left( \rho_\text{p(label)}' ( \rVector ) + \rho_\text{p(unlabel)}' ( \rVector ) \right)^2 \right.  \notag \\
	& \> \left. + \frac{1}{3}w' \left(\rho_\text{p(label)}' ( \rVector ) + \rho_\text{p(unlabel)}' ( \rVector ) \right)^3 \right),
\end{align}
\begin{equation}
	\label{eq:DimensionlessPolymerDensity}
	\begin{aligned}
		\rho_\text{p(label)}' ( \rVector ) &= \left. \rho_\text{S(label)} ( \rVector ) R_e^3 \middle/ N \right. ,  \\
		\rho_\text{p(unlabel)}' ( \rVector ) &= \left. \rho_\text{S(unlabel)} ( \rVector ) R_e^3 \middle/ N \right. ,
	\end{aligned}
\end{equation}
where $\rho_\beta ( \rVector )$ denotes the local volumetric number density of $\beta$-segments at the spatial coordinate $\rVector$. $v'$ and $w'$ are positive dimensionless constants and define the attractive and repulsive interaction strengths among the segments, respectively. $\rho_\text{p(label)}' ( \rVector )$ and $\rho_\text{p(unlabel)}' ( \rVector )$ are the dimensionless local molecular densities of the labelled and unlabelled molecules, respectively. Equation~\eqref{eq:NonidealFreeEnergyDimensionless} indicates that both the molecular species identically interact with each other, and that the system is equivalent to the single-component solvent-free model~\cite{DoctoralThesis,Norizoe:2013EPL,Norizoe:2014JCP}. This is the same as the relation between deuterated and non-deuterated polymers in experiment.

\subsection{Mean-field theory}
\label{subsec:Mean-fieldTheory}
Here, the phase behaviour of bulk systems of the single-component solvent-free model is quickly summarized~\cite{DoctoralThesis,Norizoe:2013EPL,Norizoe:2014JCP}. $\rho_p' = n_p R_e^3 / V$ is the dimensionless average volumetric molecular density of the system, where $V$ denotes the system volume and $n_p$ is the number of molecules in the system. These bulk systems exhibit macrophase separation in regions of extremely large $v'$ and finite $w'$ because the molecules aggregate. This results in a dense liquid droplet of the molecules floating in a dilute gas phase. By contrast, the system stays in a homogeneous phase in regions of finite $v'$ and extremely high $w'$. The binodal line and phase diagram are constructed in $\rho_p' v'$-plane at fixed $w'$ utilizing mean-field theory~\cite{DoctoralThesis,Doi:IntroductionToPolymerPhysics}. The result of this mean-field calculation demonstrates that the parameter $v'$ corresponds to the inverse temperature, and that the quality of the implicit solvents at fixed $w'$ decreases when $v'$ rises. The critical point of this binodal line calculated based on the mean-field theory~\cite{DoctoralThesis,Norizoe:2013EPL,Norizoe:2014JCP} is denoted by $\left( \rho_p' = \rho_{pc}' = 1 / \sqrt{ 2w' } \, , \, v' = v'_c = 2 \sqrt{2w'} \, \right)$, where the suffix ``$c$'' represents this critical point. This illustrates that the binodal line vanishes at $w' = 0$, and that the macrophase separation occurs only when $w'$ is set at a non-zero value~\cite{Daoulas:2010}. $H_\text{spring}$ and $H_{\text{non-bonded}}$ correspond to the ideal and non-ideal parts of the free energy in this mean-field calculation, respectively.

\subsection{Creation of the dense fluid film}
\label{subsec:CreationOfTheDenseFluidFilm}
In regions within the binodal lines of the above phase diagram at fixed $w'$, the system is separated into two homogeneous disordered phases; the high-density liquid and dilute gas. This ``homogeneous'' dense liquid domain floating in the dilute gas prefers reducing the surface area. Therefore, when this liquid is, in the initial state, arranged in a planar layer and reaches itself through periodic boundary condition of the system box, this layer lasts for a long time because additional surface area is required when the layer ranging infinitely through the boundary of the system box turns into a sphere~\cite{Daoulas:2010,DoctoralThesis,Norizoe:2010Faraday}. This results in a homogeneous dense fluid film floating in the gas. We create and simulate this resulting homogeneous disordered film.

The parameters of the interaction strengths are fixed at $w' = 1.0$ and $v' / v'_c = 7.0$ in the present work. This is based on the previous work~\cite{DoctoralThesis} in which the phase behaviour of the bulk single-component system at this $w' = 1.0$ was investigated. With this value set of $w'$ and $v'$, the macrophase separation of the molecules occurs in the underlying single component bulk system in a region of $\rho_p' / \rho_{pc}' \sim 1$, where the simulation is performed. According to the above mean-field theory, the molecules are segregated into a dense liquid domain with the molecular density $\approx \! \! 21 \rho_{pc}'$ and a dilute gas domain in the present simulation system.

\subsection{Computation}
\label{subsec:Computation}
Note that, unlike conventional systems composed of particles interacting \textit{via} pair potential such as Lennard-Jones potential, the non-bonded interaction and local molecular densities, \textit{i.e.} equations~\eqref{eq:NonidealFreeEnergyDimensionless} and \eqref{eq:DimensionlessPolymerDensity}, are calculated using a collocation lattice in the present simulation~\cite{Mueller:2009PCCP,Norizoe:2010Faraday,Norizoe:2013EPL,Norizoe:2014JCP}. For this calculation, the system rectangular cuboid box is partitioned in a cubic lattice with the grid spacing $\varDelta L$. The centres of the coarse-grained segments can overlap with each other and they interact \textit{via} such soft potential instead of the conventional pair potentials containing a harsh excluded volume. The absence of this harsh excluded volume in the interaction potential, which is similar to dissipative particle dynamics (DPD) simulation method ~\cite{Groot:1997,Groot:1998}, significantly facilitates Monte Carlo moves and reduces relaxation time of the system to equilibrium. Furthermore, the calculation of the non-bonded interaction based on the collocation lattice is computationally advantageous because each coarse-grained segment in our model system interacts with a large number of neighbours.

The non-bonded interaction of the present solvent-free model is defined as the third-order virial expansion functional. One can choose the fourth or higher-order expansion instead of the present third-order expansion, whereas this raises the computational cost and complicates the physical background of the system. The film can be created and simulated using the present third-order virial expansion, so that this is chosen in the present work.

\section{Simulation methods}
\label{sec:SimulationMethods}
Molecular Monte Carlo simulations are performed with the canonical ensemble in 3-D, using the standard Metropolis algorithm~\cite{ComputerSimulationOfLiquids,Frenkel:UnderstandingMolecularSimulation2002}. The Mersenne Twister algorithm is utilized as a random number generator~\cite{MersenneTwister1,MersenneTwister2,MersenneTwister3}.
$L_\alpha$ denotes the size of the rectangular cuboid of the system box, where $\alpha$ denotes the Cartesian axes $x, y$, and $z$. This system box lies in regions of $0 \le \alpha < L_\alpha$. A periodic boundary condition is applied to the system. The large thin hard planar substrate is placed on $xy$-plane. $L_z = 6.0 R_e$ and $N = 32$ are fixed. $\varDelta L = (1/6) R_e$ is fixed and defines the spatial range of the non-bonded interaction between the segments~\cite{Norizoe:2013EPL}. In one simulation step, a particle (segment) is randomly selected and given a uniform random trial displacement within a cube of edge length $2 \varDelta L$. One Monte Carlo step (MCS) is defined as $n_S^\text{(total)}$ simulation steps, during which each particle is selected for the trial displacement once on average. $n_S^\text{(total)}$ denotes the total number of segments in the system. These simulation conditions, which are the same as our recent work on the brush~\cite{Norizoe:2014JCP}, allow us to compare the simulation results between the systems of the film and brush.

One end of each molecule is randomly distributed in the initial state in regions of $0 \le x < L_x$, $0 \le y < L_y$, and $0 < z \le \varDelta L$. Using normal random numbers, the initial conformation of each molecule, \textit{i.e.} the coordinates of the other segments, is arranged in random coils~\cite{Kawakatsu:StatisticalPhysicsOfPolymersAnIntroduction} with $\approx \! \! R_e$ of the root mean square end-to-end distance. This illustrates that the molecules are initially distributed in $0 \le x < L_x$, $0 \le y < L_y$, and $0 < z \lesssim R_e$, and that the film is constructed in this spatial region in a short time after the simulation starts.

\subsection{Homogeneous coating on the hard substrate}
\label{subsec:HomogeneousCoatingOnTheSubstrate}
\begin{figure*}[!tb]
	\centering
	\includegraphics[clip,width=15cm]{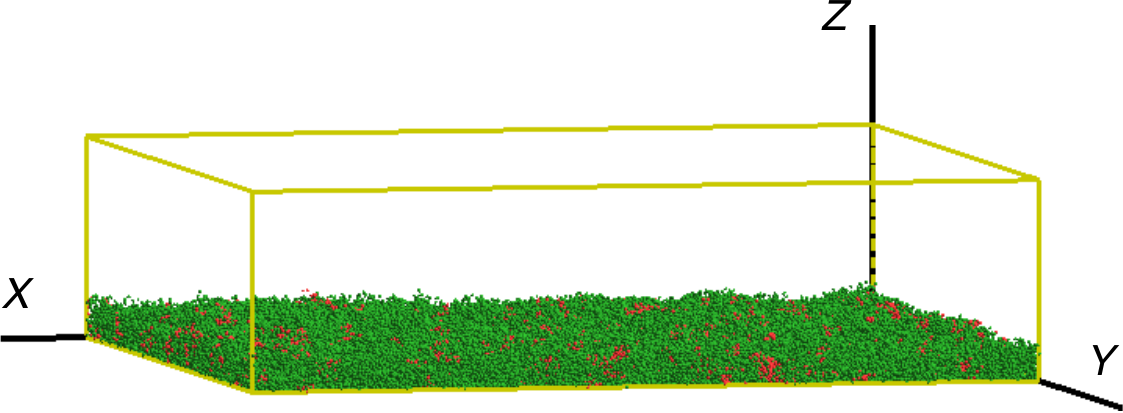}
	\caption{Snapshot of the system at $n_p^\text{(total)} = 7331$, $n_\text{p(label)} = 576$, and $( L_x = 24.0 R_e, L_y = 24.0 R_e, L_z = 6.0 R_e )$. $\rho_p^\text{(total)} R_e^3 / \rho_{pc}' = 3.00$ and $\rho_\text{p(label)}' / \rho_{pc}' = 1.41$. Red and green spheres represent S(label) and S(unlabel)-segments, respectively. Sampled at $6 \times 10^5$ MCS.}
	\label{fig:SinSolFreeFilmN32W10Lx240Ly240Lz60VVc70np7331npA576_000600000MCS}
\end{figure*}

An altitude of the created disordered fluid film is always fluctuating when the film is placed in the bulk system, which is similar to a random walk. Therefore, here we hinder the film leaving from the substrate, using monomers (polymerization degrees equal to 1) of S(unlabel)-species spatially fixed at a square lattice with the lattice const. $\varDelta L$ on the thin hard planar substrate laid at $z = 0$. This lattice const. $\varDelta L$ indicates that these fixed monomers tightly covers and continuously and homogeneously coats the substrate because $\varDelta L$ denotes the spatial range of the non-bonded interaction. The substrate itself disallows the molecules to pass through and applies no other interactions, such as friction, to the molecules. However, these fixed monomers homogeneously distributed on the planar substrate in the present simulation system attract the whole dense fluid membrane, rather than each molecule, to the substrate. Therefore, the whole dense fluid membrane covers the whole substrate. The segments of the labelled and unlabelled molecules are distributed and form a dense fluid layer in $0 < z \lesssim R_e$. An example of the constructed film is displayed in Fig.~\ref{fig:SinSolFreeFilmN32W10Lx240Ly240Lz60VVc70np7331npA576_000600000MCS}. Note that both the molecular species can freely move and diffuse inside the resulting molecular film although the whole film is fixed on and covers the substrate. When one segment moves in the lateral direction of the continuously-coated substrate covered with the homogeneous film, the interaction between this moving segment and the segments spatially fixed on the substrate is unchanged. This illustrates that no interaction in the lateral direction acts between the homogeneously-coated substrate and each segment of the film. The average volumetric density of the labelled molecules in this region of the layer is defined~\cite{Norizoe:2013EPL,Norizoe:2014JCP} as $\rho_\text{p(label)} = n_\text{p(label)} / ( L_x L_y \times R_e )$, where $n_\text{p(label)}$ denotes the number of labelled molecules. This average density is reduced to the dimensionless density $\rho_\text{p(label)}' = \rho_\text{p(label)} R_e^3$.

On the other hand, one end of each polymer is directly and spatially fixed on the substrate in the system of the brush. This constraint, \textit{i.e.} grafting, disallows the diffusion of the polymers. Free ends of the grafted polymers are allowed to move only in the vicinity of the grafting point of each polymer. This is the difference between the systems of the film and brush.

\subsection{Sampling}
\label{subsec:Sampling}
\begin{figure}[!tb]
	\centering
	\includegraphics[clip,width=7cm]{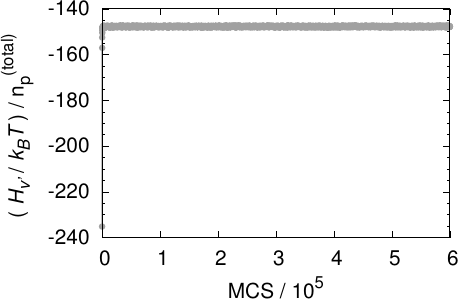}
	\caption{The time evolution of $H_{v'} / k_B T$, defined in eq.~\eqref{eq:FreeEnergy2ndVirialDimensionless}, of the system at $n_p^\text{(total)} = 7331$, $n_\text{p(label)} = 576$, and $( L_x = 24.0 R_e, L_y = 24.0 R_e, L_z = 6.0 R_e )$, which is the same as the system displayed in Fig.~\ref{fig:SinSolFreeFilmN32W10Lx240Ly240Lz60VVc70np7331npA576_000600000MCS}. $n_p^\text{(total)}$ denotes the total number of labelled and unlabelled molecules in the system.}
	\label{fig:SinSolFreeFilmN32W10Lx240Ly240Lz60VVc70np7331npA576-PotentialEnergy}
\end{figure}

Figure~\ref{fig:SinSolFreeFilmN32W10Lx240Ly240Lz60VVc70np7331npA576-PotentialEnergy} shows the time evolution of $H_{v'} / k_B T$ of the system displayed in Fig.~\ref{fig:SinSolFreeFilmN32W10Lx240Ly240Lz60VVc70np7331npA576_000600000MCS}, where $H_{v'} / k_B T$ denotes the first term of eq.~\eqref{eq:NonidealFreeEnergyDimensionless},
\begin{equation}
	\label{eq:FreeEnergy2ndVirialDimensionless}
	\frac{ H_{v'} }{k_B T} := \int_V \frac{dV}{{R_e}^3} \left( -\frac{1}{2}v' \left( \rho_\text{p(label)}' ( \rVector ) + \rho_\text{p(unlabel)}' ( \rVector ) \right)^2 \right).
\end{equation}
This result indicates that the system reaches equilibrium in a short time after the simulation starts, as discussed above. $H_{v'} / k_B T$, \textit{i.e.} eq.~\eqref{eq:FreeEnergy2ndVirialDimensionless}, is smaller than the second term of eq.~\eqref{eq:NonidealFreeEnergyDimensionless} and sensitive to the structure of the system, so that the equilibration of the system is quickly evident from the time evolution of eq.~\eqref{eq:FreeEnergy2ndVirialDimensionless}. After $1.0 \times 10^5$ MCS, by which the system relaxes to the equilibrium state, data are collected every $10^4$ MCS till $6 \times 10^5$ MCS and 51 independent samples of particle configurations are obtained in the present work.

\subsection{Four films created in the present work}
\label{subsec:FourFilmsCreatedInThePresentWork}
\begin{figure}[!tb]
	\centering
	\includegraphics[clip,width=7.5cm]{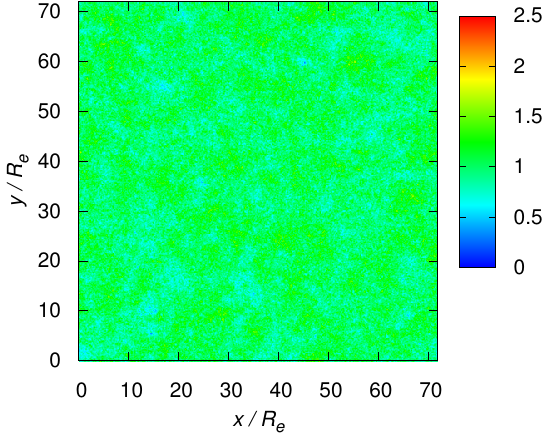}
	\caption{The lateral molecular density distribution, $\rho_p^{(\text{area})} ( x, y ) / \rho_p^{(\text{area})}$, of FilmLarge 2, where $\rho_p^{(\text{area})} = n_p^\text{(total)} / L_x L_y$ is the average areal density of both the molecular species. $\rho_p^{(\text{area})} ( x, y )$ is defined in eq.~\eqref{eq:LateralMolecularDensityOfFilm}. $n_\text{p(label)} = 5184$. Sampled at $6 \times 10^5$ MCS. This density distribution shows that the whole substrate is covered in the disordered fluid membrane.}
	\label{fig:SinSolFreeFilmN32W10Lx720Ly720Lz60VVc70np109969npA5184_000600000MCS-SpatialSegmentDistributionForColumnsZ}
\end{figure}

$n_\text{p(unlabel)}$ denotes the number of unlabelled molecules. $n_p^\text{(total)} = n_\text{p(label)} + n_\text{p(unlabel)}$ is the total number of labelled and unlabelled molecules in the system. The value of $\rho_p^\text{(total)} = n_p^\text{(total)} / V$ determines the thickness of the molecular layer when $L_z$ is fixed. In the present work, we mostly show simulation results at $( L_x = 24.0 R_e, L_y = 24.0 R_e, L_z = 6.0 R_e )$ and each value of $n_p^\text{(total)} = 7331$ (referred to as FilmSmall 1) and 12218 (FilmSmall 2), resulting in the average dimensionless molecular density $\rho_p^\text{(total)} R_e^3 / \rho_{pc}' = 3.00$ and 5.00 respectively. The systems at various values of $\rho_\text{p(label)}'$ are simulated for each of these two films, and the percolation threshold of the labelled molecules is determined. We have confirmed that the physical properties of the system are not significantly changed even when a larger system with the same values of $\rho_p^\text{(total)}$ at $( L_x = 72.0 R_e, L_y = 72.0 R_e, L_z = 6.0 R_e )$ is simulated. The films simulated in the present work are listed in Table~\ref{tab:ListOfFilms}. As an example, Fig.~\ref{fig:SinSolFreeFilmN32W10Lx720Ly720Lz60VVc70np109969npA5184_000600000MCS-SpatialSegmentDistributionForColumnsZ} shows the lateral molecular density distribution of FilmLarge 2, which demonstrates that the film covers the whole substrate, where the lateral density is defined as,
\begin{equation}
	\label{eq:LateralMolecularDensityOfFilm}
	\rho_p^{(\text{area})} ( x, y ) := \int_0^{L_z} dz \, \rho_p ( x, y, z ).
\end{equation}
$\rho_p ( x, y, z )$ denotes the total of the local volumetric densities of both the molecular species at the spatial position $( x, y, z )$. $\rho_p^{(\text{area})} ( x, y )$ has a dimension of areal density.
\begin{table}[!htb]
	\caption[]{List of the 4 films simulated in the present work: FilmSmall 1, FilmSmall 2, FilmLarge 1, and FilmLarge 2. FilmLarge 1 and FilmLarge 2 are mostly simulated for the measurement of the critical exponents. $n_S^\text{(total)} = N n_p^\text{(total)}$ is the total number of segments in the system.}
	\label{tab:ListOfFilms}
	\centering
	\begin{tabular}{cccccc} \hline
		& $L_x / R_e$ & $L_y / R_e$ & $L_z / R_e$ & $n_p^\text{(total)}$ & $\left. \rho_p^\text{(total)} R_e^3 \middle/ \rho_{pc}' \right.$  \\ \hline
		FilmSmall 1 & 24.0 & 24.0 & 6.0 & 7331 & 3.00  \\
		FilmSmall 2 & 24.0 & 24.0 & 6.0 & 12218 & 5.00  \\
		FilmLarge 1 & 72.0 & 72.0 & 6.0 & 65981 & 3.00  \\
		FilmLarge 2 & 72.0 & 72.0 & 6.0 & 109969 & 5.00  \\ \hline
	\end{tabular}
\end{table}

Our purpose is to determine and compare the critical exponents between the films and brushes in the equal condition. For this purpose, the system size of these four films as well as the simulation and data analysis methods is the same as the ones of the brushes, and finite size scaling~\cite{Stauffer1985} is not chosen in the present work because this technique was not used in the study on the brushes~\cite{Norizoe:2014JCP}.

\subsection{Definitions of cluster size and percolation cluster}
\label{subsec:DefinitionsOfClusterSizeAndPercolationCluster}
The definitions of cluster size and percolation cluster are the same as the ones for the brush~\cite{Norizoe:2014JCP}. First, a square lattice is placed on the substrate. The lattice const. of this is fixed at $\varDelta L$. Indices $( i_x, i_y )$ for $0 \le i_\alpha < L_\alpha / \varDelta L$ denote each cell of the lattice. We assume that the labelled molecules fill a cell of $( i_x, i_y )$ when one or more S(label)-segments are found in a rectangular parallelepiped region of $i_x \le x / \varDelta L < i_x + 1$,  $i_y \le y / \varDelta L < i_y + 1$, and $0 \le z < L_z$. We also assume that a pair of filled cells, denoted by $( i_x, i_y )$ and $( i'_x, i'_y )$ respectively, is linked when a relation, $\left| i_x - i'_x \right| \le 1$ and $\left| i_y - i'_y \right| \le 1$, is satisfied. A cluster with the size $M$ is defined as an isolated single network of links composed of $M$ filled cells. A large cluster bridging both the edges of the hard planar substrate is referred to as a percolation cluster. The cluster size distribution~\cite{Stauffer1985}, denoted by $n(s)$ where $s$ is the cluster size, is defined as, $n(s) = m(s) / N_\text{cell}$, where $m(s)$ is the number of clusters with the size $s$ found in the system and $N_\text{cell} = L_x L_y / ( \varDelta L )^2$ denotes the total number of the cells.

The disordered fluid membranes are simulated in the present work. The analyzed structures, which are clusters, are composed of labelled molecules, which are randomly chosen from the whole membrane prior to the start of the simulation runs. In short, the randomly chosen molecules in the disordered membrane are labelled and watched. These molecules encounter each other and create clusters, or instantaneously separate from and break the created clusters. Furthermore, the molecules of the solvent-free model can freely move and diffuse, as discussed in section~\ref{sec:Solvent-freeModel}. The clusters composed of such labelled molecules quickly appear or disappear. The cluster structures also change in a short time.

\section{Simulation results}
\label{sec:SimulationResults}
\begin{figure*}[!tb]
	\centering
	\includegraphics[clip,width=10cm]{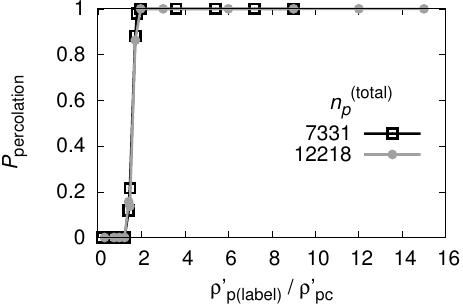}
	\caption{Occurrence probability of percolation clusters of the labelled molecules, denoted by $P_\text{percolation}$, with the system size $( L_x = 24.0 R_e, L_y = 24.0 R_e, L_z = 6.0 R_e )$ at $n_p^\text{(total)} = 7331$ and 12218, i.e. FilmSmall 1 and FilmSmall 2 respectively.}
	\label{fig:SinSolFreeFilmN32W10Lx240Ly240Lz60VVc70np7331np12218}
\end{figure*}

\begin{figure}[!tb]
	\centering
	\includegraphics[clip]{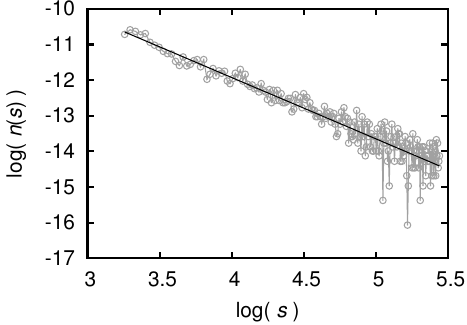}
	\caption{Cluster size distribution of the labelled molecules for FilmLarge 1 at $\rho_\text{p(label)}' / \rho_{pc}' = 1.58$, at which the percolation transition occurs. $\rho_p^\text{(total)} R_e^3 / \rho_{pc}' = 3.00$, the same as FilmSmall 1. $n_\text{p(label)} = 5800$. $\log (s)$-$\log (n(s))$ graph is plotted, where ``$\log$'' denotes the natural logarithm. The black line represents a linear fitting of this graph, $y = -\tau * x + b$ with $\tau = 1.7$ and $b = -5.04$.}
	\label{fig:SinSolFreeFilmN32W10Lx720Ly720Lz60VVc70np65981npA5800-ClusterSizeDistributionForZeroDen}
\end{figure}

Here the simulation results are discussed. Occurrence probability of the percolation clusters of the labelled molecules inside the film at each value of $\rho_\text{p(label)}'$ is presented in Fig.~\ref{fig:SinSolFreeFilmN32W10Lx240Ly240Lz60VVc70np7331np12218} for FilmSmall 1 and FilmSmall 2. When $\rho_\text{p(label)}'$ is raised from zero, this occurrence probability abruptly rises from 0 to 1 at $\rho_\text{p(label)}' / \rho_{pc}' \approx 1.5$. This boundary sharply divides regions of percolation and non-percolation phases. This result is consistent with the sharp boundary in the occurrence probability at the percolation transition point in usual percolation phenomena~\cite{Stauffer1985,Norizoe:2014JCP}. The cluster size distribution, $n(s)$, is measured in the vicinity of this sharp boundary. The results at $( L_x = 72.0 R_e, L_y = 72.0 R_e, L_z = 6.0 R_e )$ and $n_p^\text{(total)} = 65981$ (referred to as FilmLarge 1) and 109969 (FilmLarge 2), which are consistent with the values of $\rho_p^\text{(total)}$ of FilmSmall 1 and FilmSmall 2 respectively, indicate a relation, $n(s) \propto s^{-\tau}$ with $\tau \approx 1.7 \pm 0.1$, independently of the values of $\rho_p^\text{(total)}$ and $\rho_\text{p(label)}'$. An example of this result is shown in Fig.~\ref{fig:SinSolFreeFilmN32W10Lx720Ly720Lz60VVc70np65981npA5800-ClusterSizeDistributionForZeroDen}. The power law of $n(s)$ is also universally observed in usual percolation phenomena in the vicinity of the percolation transition point~\cite{Stauffer1985,Norizoe:2014JCP}. These show that the sharp boundary is the percolation threshold of the labelled molecules. $\tau$ is a critical exponent and referred to as Fisher exponent. The present result, $\tau \approx 1.7$, is similar to $\tau \approx 1.6$ of the 3-D system of the brush~\cite{Norizoe:2014JCP} and smaller than $\tau = 1.9$ and 2.2 of a model colloidal system in 2 and 3 dimensions respectively~\cite{Norizoe:2005,Norizoe:2012JCP}. These results indicate that the values of the percolation threshold and $\tau$ are not significantly dependent on the film thickness in the region of the thickness $\approx \! \! R_e$ because the system is projected onto the 2-dimensional planar substrate when the percolation phenomena are analyzed.
\begin{figure}[!tb]
	\centering
	\includegraphics[clip]{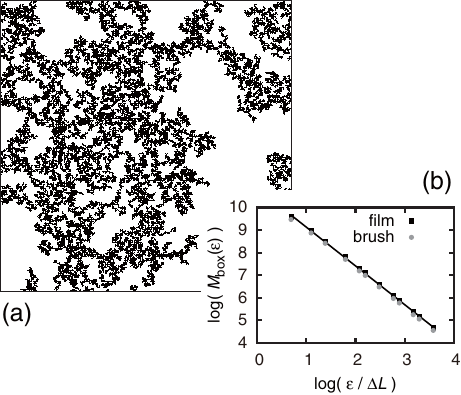}
	\caption{(a) An example of a snapshot of the percolation cluster (large cluster) of the labelled molecules for FilmLarge 1 at $\rho_\text{p(label)}' / \rho_{pc}' = 1.58$, at which the percolation transition occurs. $n_\text{p(label)} = 5800$. Sampled at $31 \times 10^4$ MCS.
		(b) A result of box counting of the percolation clusters for the same film (FilmLarge 1) and $\rho_\text{p(label)}' / \rho_{pc}'$ as (a). $\log (\epsilon)$-$\log (M_\text{box}(\epsilon))$ graph is plotted, where $\epsilon$ denotes the box size and $M_\text{box}(\epsilon)$ is the number of covered boxes with the size $\epsilon$ found in the system. ``$\log$'' denotes the natural logarithm. Black points show the result. A solid black line represents a linear fitting of the graph, which is $y = -D_f * x + b$ with $D_f = 1.7$ and $b = 10.84$. $D_f$ is fractal dimension or box counting dimension. For reference, the result of the brush is also shown by grey points (data copied from Ref.~\cite{Norizoe:2014JCP}).}
	\label{fig:SinSolFreeFilmN32W10Lx720Ly720Lz60VVc70np65981npA5800_000310000MCS-ConfPercolatedForZeroDen}
\end{figure}

The percolation threshold, \textit{i.e.} transition point, is found in regions of $\rho_\text{p(label)}' = \rho_\text{p(label)} R_e^3 \sim 1$, because the coils of the labelled molecules, whose size $\approx \! \! R_e$, start to overlap with each other in the vicinity of this value of $\rho_\text{p(label)}'$ when $\rho_\text{p(label)}'$ rises from zero. This result of the present system is consistent with that of the polymer brush.

Next, structure of the percolation cluster in the vicinity of the percolation threshold is studied. A snapshot of the percolation cluster at $\rho_\text{p(label)}' / \rho_{pc}' = 1.58$ for FilmLarge 1, at which the percolation transition occurs, is shown in Fig.~\ref{fig:SinSolFreeFilmN32W10Lx720Ly720Lz60VVc70np65981npA5800_000310000MCS-ConfPercolatedForZeroDen}(a) as an example. A complicated pattern is observed in this snapshot. The fractal dimension (box counting dimension) of the percolation clusters of the labelled molecules, denoted by $D_f$, is measured using box counting method in the vicinity of the percolation transition point for FilmLarge 1 and FilmLarge 2. The results of this measurement indicate $D_f \approx 1.7 \pm 0.1$ in the vicinity of the transition point, independently of the values of $\rho_p^\text{(total)}$ and $\rho_\text{p(label)}'$. An example of these results is given in Fig.~\ref{fig:SinSolFreeFilmN32W10Lx720Ly720Lz60VVc70np65981npA5800_000310000MCS-ConfPercolatedForZeroDen}(b). $D_f = 1.7$ is quantitatively consistent with $D_f = 1.7$ of the system of the polymer brush~\cite{Norizoe:2014JCP} and qualitatively consistent with $D_f = 91 / 48 = 1.9$ of conventional 2-D lattice models of percolation~\cite{Stauffer1985}.

The above simulation results demonstrate that the percolation transition occurs in the dynamic networks, and that the universality class is independent of the extra-molecular topological constraints.

\subsection{Universality of the percolation threshold}
\label{subsec:UniversalityOfThePercolationthreshold}
The percolation transition occurs at the volume fraction of the particles $\approx \! \! 0.4$ in 2 dimensions (2-D) and $\approx \! \! 0.16$ in 3-D in the conventional percolation systems~\cite{Zallen:PhysicsOfAmorphousSolids}. These values are dependent only on the dimensionality, and independent from characteristics of each system, such as the lattice structure~\cite{Zallen:PhysicsOfAmorphousSolids}. Here we examine the validity of this universal phase behaviour in the present system.

The collocation grid defined in section~\ref{sec:Solvent-freeModel} partition the system box into
small cubes of edge length $\varDelta L$. We assume that the labelled molecules fill a small cube when one or more S(label)-segments are found in this small cube. The volume fraction of the molecules in 3-D in the present system is defined as the ratio of the number of filled small cubes to the total number of small cubes.

The transition density of the present system is found in regions of $\rho_\text{p(label)}' = \rho_\text{p(label)} R_e^3 \sim 1$, as discussed above. At this density of $\rho_\text{p(label)} R_e^3 \sim 1$, $N$ segments of S(label)-species, \textit{i.e.} one labelled molecule, are, on average, found in the cubic unit volume of $R_e \times R_e \times R_e = R_e^3$ inside the film. The collocation grid defined in section~\ref{sec:Solvent-freeModel} partitions this cubic space into $(R_e / \varDelta L)^3$ small cubes. Here we homogeneously divide the labelled molecule into $n_\text{sub} = \left. N \middle/ N_\text{sub} \right.$ subchains, where $N_\text{sub}$ denotes the number of segments in one subchain, and assume that the root mean square of the end to end distance of each subchain is set at $\varDelta L$. This results in relations,
\begin{gather}
	\label{eq:EndToEndDistanceOfSubchains}
	\varDelta L = b \sqrt{ N_\text{sub} } = R_e \sqrt{ N_\text{sub} / N },  \\
	\label{eq:NumberOfSubchains}
	n_\text{sub} = N / N_\text{sub} = ( R_e / \varDelta L )^2,
\end{gather}
where $b$ denotes the average bond length between bonded segments of ideal molecular chains,
\begin{equation}
	\label{eq:AverageBondLength}
	b = \left. R_e \middle/ \sqrt{N} \right..
\end{equation}
The labelled molecule fills $\approx \! \! n_\text{sub}$ small cubes of the $(R_e / \varDelta L)^3$ small cubes because each subchain fills $\approx \! \! 1$ small cubes according to the end to end distance of the subchains, $\varDelta L$. Therefore, the volume fraction is, using eq.~\eqref{eq:NumberOfSubchains}, equal to,
\begin{equation}
\label{eq:VolumeFractionAtTransitionPresentSystem}
	\frac{ n_\text{sub} }{ (R_e / \varDelta L)^3 } = \frac{ \varDelta L }{ R_e }.
\end{equation}
In the present system, this equals 1/6, which is consistent with the above universal value, $\approx \! \! 0.16$, of the 3-D conventional percolation systems. This demonstrates the compatibility of the percolation threshold between the conventional and present systems. A similar result also holds for the systems of the grafted molecules, \textit{i.e.} the brushes. This compatibility is independent from characteristics of each system, \textit{i.e.} the extra-molecular topological constraint (grafting). In other words, these results indicate that the above universal phase behaviour of the conventional percolation systems also applies to the molecular systems independently from the grafting.

The volume fraction at the transition in the present system discussed above, eq.~\eqref{eq:VolumeFractionAtTransitionPresentSystem}, depends explicitly on the grid spacing, $\varDelta L$. For example, when $\varDelta L / R_e$ decreases from the present value 1/6, this volume fraction could also decrease. This could contradict the above universality of the volume fraction of the particles at the percolation transition. However, the decrease in $\varDelta L$ results in the decrease in the volume of each small cube, \textit{i.e.} the spatial region which each segment fills. The probability that a pair of adjacent labelled molecules overlaps also decreases. This raises the transition density from the present value of $\rho_\text{p(label)} R_e^3 \sim 1$. In other words, at percolation transition at low $\varDelta L$, the number of filled small cubes increases although the volume of each small cube decreases. This suggests that the value of the volume fraction at the transition could be kept at the universal value, whereas the universality of the volume fraction for the values of $\varDelta L$ is beyond the scope of the present work.

When the system is projected onto the 2-D planar substrate to analyze the percolation phenomena, the volume fraction in 2-D is defined as the ratio of the number of filled cells to the total number of cells. This volume fraction is measured in the vicinity of the percolation transition point for the four simulated films in the present work, and the result is $\approx 0.4 \pm 0.05$ for all the films. The above compatibility of the percolation threshold between the conventional and present systems also holds for the 2-D case as well as 3-D.

\subsubsection{Total volume fraction inside the film}
\label{subsubsec:TotalVolumeFractionInsideTheFilm}
The labelled and unlabelled molecules fill the spacial region of the molecular layer and compose the dense fluid membrane. The small cubes in the molecular layer are filled with these molecules. In this sense, the total volume fraction of both the molecular species inside the film $\approx \! \! 1$.

\subsection{Universality and molecular conformation}
\label{subsec:UniversalityAndMolecularConformation}
The percolation transition occurs at the density where the molecules start to overlap with each other, as discussed above. At this density, the molecular conformation of the grafted molecule is similar to that of the isolated grafted molecule, which is also similar to that of the non-grafted molecule in the film. Furthermore, the number of overlaps between the molecules is $\sim \! \! 0$, so that the network of the percolation clusters is composed of the apposed, and partially overlapping with the neighbours, molecules. This illustrates that the composing element of the percolation clusters, \textit{i.e.} molecular conformation of each molecule, is similar between the films and brushes. Therefore, the fractal dimension, \textit{i.e.} self-similarity, of the percolation clusters at the transition point also show the similar values between the films and brushes. A similar result also holds for the compatibility of the cluster size distribution at the transition point between these systems.

Thus, the universality of the percolation transition and concomitant critical phenomena is independent from the extra-molecular topological constraints, \textit{i.e.} grafting. In other words, both the systems of the brush and film show the same values of the critical exponents because these systems are the same except that the polymers of the brushes are grafted. This universality also holds for systems at different value sets of simulation parameters, such as $w'$ and $v'$, as long as the molecular conformation is similar.

\subsection{Spatial dimension of the system}
\label{subsec:SpatialDimensionOfTheSystem}
The present results, $\tau \approx 1.7$ and $D_f \approx 1.7$, and scaling relation~\cite{Stauffer1985} result in,
\begin{equation}
\label{eq:ScalingRelationSpatialDimension}
	d = D_f ( \tau - 1 ) \approx 1.19,
\end{equation}
where $d$ denotes the spatial dimension of the system. This value of $d$ is different from both 2 and 3. The present system is simulated in 3-D, and mapped onto 2-D for the data analysis of the percolation phenomena. The mismatch, $d \ne 2 \text{ and } 3$, could be attributed to this intermediate and/or multiple dimensionality of the simulation system, whereas this mismatch is beyond the scope of the present work.

\section{Conclusions}
\label{sec:Conclusions}
In conclusion, the percolation and critical phenomena of the long flexible mobile molecules in dense fluid films on planar hard substrates have been simulated using the molecular Monte Carlo method in a coarse-grained scale in 3-D. The percolation threshold (transition density) has been determined. Fisher exponent of the system, $\tau \approx 1.7$, has also been determined. Fractal structure of the percolation clusters at the percolation threshold has been revealed. $D_f \approx 1.7$. These values of $\tau$ and $D_f$ are consistent with the results of the polymer brush. This demonstrates that the structure of the dynamic network among the molecules in the vicinity of the percolation transition point is independent of the grafting of the polymers, \textit{i.e.} topological constraints. Furthermore, the percolation threshold is also independent of the topological constraints. These results provide suggestive evidence that both the universality class and percolation threshold are independent from the extra-molecular topological constraints in arbitrary physical systems and films. On the other hand, the value sets of $( \tau, D_f )$ could be dependent on the physical and chemical characteristics of each system.

\begin{acknowledgments}
	The authors wish to thank Prof. Marcus M{\"u}ller and Dr Kostas Ch. Daoulas for helpful suggestions and discussions. This work is based on results obtained from a project commissioned by New Energy and Industrial Technology Development Organization (NEDO), Japan.
\end{acknowledgments}

\section*{Author contribution statement}
Morita began and managed this research project. Under the supervision of Morita, Norizoe designed and ran the simulation, and worked on physical understanding of the simulation results. The manuscript was written by Norizoe, whereas Morita also helped revising and making improvements to the manuscript.

%

\end{document}